# Microscopic aspects of artificial ageing in Al-Mg-Si alloys

Mazen Madanat [a], Meng Liu [a,b,*], Xingpu Zhang [a], Qianning Guo [a], Jakub Čížek [c] and John Banhart [a,b]

[a] *Technische Universität Berlin, Hardenbergstraße 36, D-10623 Berlin, Germany*
[b] *Helmholtz Zentrum Berlin, Hahn-Meitner-Platz 1, D-14109 Berlin, Germany*
[c] *Charles University, V Holešovičkách 2, 18000 Praha 8, Czech Republic*

*corresponding author: meng.liu@helmholtz-berlin.de, now at Chinalco Materials Application Research Institute Co., Ltd., Beijing, China

Al-Mg-Si alloys with total solute contents ranging from 0.8 to 1.4 wt.% were solutionised, quenched and then artificially aged (AA) at 180 °C, after which positron annihilation lifetime spectroscopy was applied to obtain information about precipitation and vacancy evolution during preceding ageing. Hardness and electrical resistivity measurements were carried out to complement these measurements. AA was carried out in four different heating media, which allowed for varying the heating rate from 2.4 K·s$^{-1}$ to 170 K·s$^{-1}$. The main result of the study is that there is a competition between vacancy losses and precipitation. Any precipitation taking place during quenching or during heating to the AA temperature helps to prevent vacancies from going to sinks and allows them to assist in solute clustering. Higher solute content, slower heating to 180 °C and natural pre-ageing before AA were found to have a comparable effect.

**Frequently used abbreviations**

| | |
|---|---|
| SHT | solution heat treatment |
| NA / NPA / NSA: | natural ageing / natural pre-ageing / natural secondary ageing |
| AA | artificial ageing |
| LM | liquid metal |
| MS | molten salt |

I. INTRODUCTION

The technologically important 6XXX series of age-hardenable alloys is based on the ternary system Al-Mg-Si. After solutionising and quenching, artificial ageing (AA) at typically 180 °C leads to the formation of a series of metastable precipitates that increase strength. 6XXX alloys are well investigated and the precipitation sequence largely known. However, understanding multi-stage heat treatments or subtle influences of alloy composition is still difficult [1]. Even the simplest case – direct AA after quenching – bears some unknown problems. The earliest precipitates, also called 'atom clusters', are assumed to have the fcc structure of the host metal and to transform later to monoclinic GP zones and $\beta''$ precipitates [2-4], but how this happens and how precipitates grow is not known with certainty [5,6]. Moreover, the role of vacancies in the first stages of AA is not clear. After quenching, the high site fraction of vacancies formed at the solutionising temperature is partially preserved. During AA, excess vacancies anneal out and eventually the vacancy site fraction approaches the equilibrium value at the AA temperature, but how fast this happens is not clear because vacancy loss is influenced by both vacancy sinks such as dislocation jogs and grain boundaries [7] and by growing clusters. Solute supersaturation is also a crucial factor since it provides the driving force for precipitation. How it varies during AA is not exactly known.

We intend to contribute to the understanding of the role of vacancies and solutes especially in the early stages of AA. We study very short AA treatments down to 0.3 s and also vary the heating rate from 'room temperature' to 180 °C. The main tool employed is positron annihilation lifetime spectroscopy. This method is sensitive to both vacancies and atom clusters/precipitates and has been applied previously to explain the NA behaviour in identical or similar alloys [8-10]. The high sensitivity to vacancies is actually unique and even small changes in vacancy fraction can be detected [11]. Clusters can also be detected by atom probe tomography but as the present study involves very early stages of clustering any quantitative analysis is very challenging or even not possible [12]. X-ray small-angle scattering would fail due to insufficient elemental contrast between Al, Mg and Si [13]. Only electrical resistivity measurement could provide evidence on changes of both vacancy and clusters, which is why it is applied to provide some complementary data.[†]

We study pure ternary model alloys to keep out possible effects of both common solute atoms such as Cu that might influence precipitation and intermetallic-forming elements such as Fe and Mn that might lower the Si content. Most investigations are carried out on a lean alloy containing ~0.4 wt.% of both Mg and Si, which is similar to the industrial alloy 6060, and an alloy higher in solute containing 0.6 wt.% Mg and 0.8 wt.% Si, which is similar to the industrial alloy 6005 (all in wt.%). One objective is to specify the pronounced differences in ageing behaviour between lean Al-Mg-Si alloys (Mg+Si≤1%) and more concentrated ones.

## II. EXPERIMENTAL

### A. Samples and heat treatment

Pure ternary aluminium alloys were provided by Hydro Aluminium, Bonn, as described elsewhere [8,13]. Compositions are specified in the format 'n-m', where 'n' stands for the Mg, 'm' for the Si content in 1/10 wt.% throughout this paper, see Table 1.

Samples of $10 \times 10 \times 1$ mm$^3$ size were prepared for both positron lifetime and hardness measurements. Samples were ground and ultrasonically cleaned with alcohol for clean surfaces. Solution heat treatment (SHT) was performed at 540 °C for 1 h in argon atmosphere followed by ice water quenching. After quenching, one of the following heat treatments was carried out: (i) Natural ageing (NA) at 'room temperature' (i.e. 20 ± 2 °C). As the activation energy for NA is of the order of 80 kJ/mol [8], variations within this range can lead to changes of the NA kinetics of ~35%. (ii) Artificial ageing (AA) in one of four different media (see below) followed by measurement at ~20 °C. (iii) Natural pre-ageing (NPA) for 5, 30 or 80 min followed by AA, after which measurements were conducted at ~20 °C again.

Four different AA media held at 180 °C were used to study the influence of the heating rate to 180 °C: (i) Highly thermally conductive and low-melting liquid metal ('LM') Bi57Sn43. The Al-Mg-Si samples were kept in motion to ensure faster heating to 180 °C. After immersion into the metal the sample surfaces were checked for possible Bi-Sn residues but were found not to be wetted by the Bi-Sn alloy. The positron lifetime in Bi57Sn43 was measured to be 240 ps. No such lifetime contribution was found in any spectrum. (ii) Silicon oil ('oil') as widely used as AA medium. Samples were also kept in motion during AA. (iii) A molten salt ('MS') mixture of 53% $KNO_3$ + 40% $NaNO_2$ + 7% $NaNO_3$ with a low melting point of 142 °C [16]. During AA, the samples were kept in a resting position to obtain a slower heating rate. (iv) Controlled heating plates ('HP') that heat linearly at slower rates. Samples were sandwiched between two aluminium buffer plates to damp temperature oscillations caused by the temperature controller.

For all heating modes, A 0.5-mm thick thermocouple was inserted into holes drilled into the edge of test specimens to measure the temperature courses given in Figure 1. The time to reach 170 °C is 0.9 s for LM, which we call 'fast heating', 6 s for oil and 10 s for MS, which we call 'moderate heating' and 63 s for the heating plate, 'slow heating'. The corresponding average heating rates up to 170 °C are 170 $K·s^{-1}$, 25.5 $K·s^{-1}$, 15.3 $K·s^{-1}$ and 2.4 $K·s^{-1}$, respectively.

**B.   Positron lifetime measurements**

Most positron lifetime measurements were carried out at the laboratory in Berlin ('B'). Two identical samples were assembled to a sandwich with the positron source in between and an aluminium foil wrapped around it. This was done either after quenching from SHT or after AA and required about 2 min during which the sample remained at ~20 °C. Measurement details are given in supplementary Sec. S1. Positron annihilation spectra were then measured continuously at ~20 °C. Either a one-component lifetime $\tau_{1C}$ was obtained, or decompositions into two components labelled by numbers $\tau_i$ in the order of increasing lifetime and the corresponding intensities $I_i$. As it has been found that the higher lifetime component contains two components [9] we will adopt the notation $\tau_1$ and $\tau_{2+3}$ for the two components even if they cannot be resolved in a specific case, see discussion part. The quantities $I_i$ are fitting parameters that have to be converted to physical properties using a model such as the trapping model, see Sec. 4.1.1. Spectra were also measured at Charles University in Prague ('P') at −40 °C using a spectrometer with a higher time resolution [17] in order to slow down ageing kinetics and to be able to collect >5 × 10$^6$ annihilation events in positron spectra for a reliable two-component analysis.

## C. Hardness and electrical resistivity measurement

Hardness tests were carried out after a given AA time using a Vickers microhardness tester MHT-10. The load force was 1 N and the dwell time 10 s. 10 indentations were averaged for each sample. Electrical resistivity was measured on 400-mm long and 0.82-mm thick coiled wires using a four-point setup as described earlier [15]. After solutionising, quenching and brief drying the coil was immersed into liquid nitrogen and the resistivity was measured there. Then the coil was dipped into either LM or oil at 180 °C for 10 s, then quenched in water, after which resistivity was measured in liquid nitrogen again.

## III. RESULTS

### A. Positron lifetime measurements

#### a. Artificial ageing of lean alloy 4-4

*One-component positron lifetime.* The evolution of $\tau_{1C}$ versus AA time after heating in different media is shown in Figure 2a. The black point corresponds to the 'as-quenched' state captured ~2 min after quenching. Measurements after AA were carried out at ~20 °C, during which $\tau_{1C}$ evolved due to natural secondary ageing (NSA). Supplemental Sec. S2 explains how values corresponding to zero NSA are determined.

The course of $\tau_{1C}$ depends on the medium used for AA. Fast heating in LM causes $\tau_{1C}$ values as low as ~163 ps. This is a reduction by 80 ps with respect to the value measured directly after quenching. For 1 s AA, the values vary in the 11 experiments carried out by manually dipping a sample into the metal. $\tau_{1C}$ does not change much up to 5 min of AA, with a slight minimum for ~3 s AA in LM. Higher $\tau_{1C}$ values were measured after moderate-rate heating in oil, namely 218 ps and 197 ps after 2 s and 10 s, respectively, with an ensuing decrease to 185 ps for 1 min of AA. The $\tau_{1C}$ decrease caused by AA in oil is therefore much lower than for fast heating. AA in MS or on the heating plate for up to a few minutes yields a similar course but even slightly higher minimum values. In all heating media, $\tau_{1C}$ starts to increase for longer AA and $\tau_{1C}$ ~235 ps is found after long AA in oil.

*Two-component positron lifetimes.* In some cases it was possible to decompose spectra into two contributions $(\tau_1, \tau_{2+3})$ and the corresponding intensities $(I_1, I_{2+3})$, see Figure 2b,c. Each value represents a state after AA and is based on data accumulated during NSA for 15 to 30 min, during which $\tau_{1C}$ remains constant or is only slowly changing. The highest heating rate in LM leads to a different course than moderate and slower heating in oil, molten salt or on the heating plate. 1 s in LM produces a $\tau_{2+3}$~213 ps and an intensity $I_{2+3}$ of 54%. The high-precision measurement after 5 s AA in LM yields $\tau_{2+3}$ = 215 ± 1 ps, $I_{2+3}$ = 50 ± 1% at −40 °C, see crossed squares. After up to 5 min of AA in LM, these values do not change notably, but after 30 min of AA $I_{2+3}$ increases to 68%. After even longer AA, $\tau_{2+3}$ also increases and reaches 235 ps while $I_{2+3}$ goes up to 75%. Slower heating either in oil, MS or on the heating plate yield $I_{2+3}$ ~80% all through AA (Figure 2c), while the measured $\tau_{2+3}$ values are about the same as for AA in LM. Longer AA times (≥ 200 min) let $\tau_{2+3}$ increase to 235 ps. Lines and arrows in Figure 2b,c underline this description.

b. Artificial ageing of concentrated alloy 6-8

*One-component positron lifetime.* Figure 3 shows $\tau_{1C}$ of alloy 6-8 measured after AA in LM and oil. 1 s of AA in LM decreases $\tau_{1C}$ by ~40 ps compared to the as-quenched state, while for 2 s AA in oil, $\tau_{1C}$ is reduced only by ~10 ps. Increasing the AA time leads to a continuous increase to 213 ps after 2 min AA in LM followed by a slight decrease to a minimum of 211 ps after 30 min AA, while a continuous decrease to a minimum of 209 ps after 30 min AA in oil is observed. Even longer AA times cause an increase of $\tau_{1C}$ up to ~235 ps as in alloy 4-4.

An extra experiment was carried out by dipping a sample into LM held at 300 °C for ~0.3 s in order to achieve an even higher heating rate than in LM at 180 °C. The end temperature in this experiment was estimated 200 °C to 230 °C. $\tau_{1C}$ after this treatment is even lower than after AA for 1 s at 180 °C in LM (red point in Figure 3).

*Two-component positron lifetimes.* Reliable lifetime decompositions in alloy 6-8 are more difficult than in alloy 4-4 and only after AA for 0.3 s in LM at 300 °C a clear indication for two-component behaviour was found with a trap contribution of $\tau_{2+3}$ = 84%.

c. Further experiments

Figure 4 shows $\tau_{1C}$ after short AA for 1 s in LM for four more alloys, namely 5-5, 4-10, 8-6, and 10-4, see Table 1. The value obtained for alloy 5-5 lies slightly above that for alloy 4-4, while the values for the alloys with a combined Mg and Si content of 1.4 wt.% are much higher and lie close together. Figure 4 also presents $\tau_{1C}$ in alloy 4-4 after a combination of a NPA step of either 5, 30, or 80 min, see upper axis, and short AA for 1 s in LM. NPA does have a strong effect on $\tau_{1C}$ after AA as it increases it by up to 25 ps.

B. **Hardness**

a. Artificial ageing of lean alloy 4-4

Figure 5 shows the evolution of hardness in alloy 4-4 during AA in oil and compares it to $\tau_{1C}$. A negligible initial change in hardness up to 2 min of AA is followed by an increase to a maximum of 70 HV after ~1000 min of AA, after which over-ageing commences. The main increase in hardness takes place after 200 min of AA, which is later than the increase of $\tau_{1C}$ that sets in already after 5 min of AA. Over-ageing reduces hardness, but not $\tau_{1C}$.

b. Artificial ageing of concentrated alloy 6-8

The hardening response of alloy 6-8 shows a negligible change during the first 2 min of AA in oil, followed by a slight increase up to 5 min of AA. The further increase of hardness to 93 HV coincides with a shallow minimum of $\tau_{1C}$ after 30 min of AA. A maximum of hardness was achieved after 100 min AA, followed by a plateau ranging between 100 and 110 HV up to ~15 h of AA. Hardness therefore peaks earlier than $\tau_{1C}$, which takes 3.5 d to peak. Over-ageing for long AA times is pronounced. Figure 5 also shows some hardness values of samples AA in LM (open symbols), which point at a negligible influence of the AA medium.

## C. Electrical resistivity

Electrical resistivity changes during 10 s of AA in LM and in oil are shown in Figure 6 (as a function of $\tau_{1C}$, see discussion). Except for alloy 6-8 treated in oil, the electrical resistivity decreases during such short AA. The effects are small (ranging from -1% to +1%) and there is some experimental scatter, probably caused by variations of the manually conducted experiments including transfer between various heat baths during which some uncontrolled temperature changes and some mechanical deformation of the coil cannot be excluded. There is a clear trend that fast heating in the lean alloy reduces resistivity, whereas slow heating in the concentrated one increases it.

## IV. DISCUSSION

Positron lifetime spectra in complex systems such as Al-Mg-Si alloys containing vacancies and different types of precipitates are rarely caused by annihilation in just one kind of trap. Instead, competitive annihilation in different ways is likely: (i) Annihilation in the free bulk without any trapping. If all positrons do this, their corresponding lifetime will be around 160 ps as we measured on well annealed Al and as it is reported in the literature [18,19]. If some positrons annihilate in traps, an apparent reduced bulk lifetime ($\tau_1 < 160$ ps) is found [11]. (ii) Positron trapping and annihilation in mono-vacancies with a lifetime of 240 ps to 250 ps [18,19] or even lower. If a vacancy forms a complex with either a Mg or Si atom it has been estimated that the lifetime is modified only marginally (1-3 ps increase for Mg, 1-2 ps decrease for Si) [8,20,21]. For the case that vacancies are surrounded by more Mg and Si atoms (vacancy-cluster complex), there are no reliable data for the corresponding lifetime. (iii) Vacancy-free clusters of Mg and Si atoms, with possibly some Al in between, can also trap positrons. Lifetimes of 200 ps [8], 214 ps [22] or 215 ps [9] have been used.

Ordered coherent clusters (or 'GP zones') and β" precipitates are expected to give rise to positron lifetimes slightly lower than the disordered clusters formed during NA, e.g. 208 ps to 210 ps [22]. As precipitates become semi- or incoherent during prolonged ageing, the associated positron lifetime increases. Values above 250 ps have been suspected [23]. As the use of the terms 'cluster', 'GP zone' or '(early) precipitate' might create confusion we shall use the term 'cluster' for any small initial aggregation of atoms in the following.

In this paper, we assume that positrons annihilate either in the bulk (lifetime $\tau_1 \leq 160$ ps), in defects related to a vacancy ($\tau_3 = 245$ ps) [19], or in vacancy-free solute clusters or β" precipitates where the lifetime $\tau_2$ initially ranges from typically 210 to 215 ps [22] and eventually ≥250 ps once they have evolved into β'. We neglect annihilation at grain boundaries and dislocations, that clusters might have a range of different sizes and compositions (which in turn would lead to a range of positron lifetimes) and that the positron lifetime in vacancies is modified by the site occupation around them. Currently, there is no way to measure or calculate such lifetime distributions. In many cases, we measure a one-component positron lifetime $\tau_{1C}$ only and have to deduce indirectly what might have changed its value during ageing. As positrons can annihilate in three different ways, changes of the average of the three partial lifetimes can be explained by an increase or decrease of one or more of the three contributions as outlined in Figure 7.

The evolution of $\tau_{1C}$ during NA has been studied previously [8-10] and is also given in supplementary Fig. S1 (full symbols). The first $\tau_{1C}$ value measured after quenching is 243 ps in alloy 4-4 and 233 ps in alloy 6-8. Previous work has shown that in alloy 4-4, at least 85% of the positrons annihilate in vacancy-related traps in this as-quenched state [9]. The initial value in alloy 6-8 is 10 ps lower in accordance with Ref. [8]. Possibly, during quenching clustering sets in and the corresponding positron trapping component with a typical lifetime ~215 ps reduces $\tau_{1C}$. The extent of such clustering is difficult to quantify since clustering and vacancy losses both lower the average positron lifetime. Temperature-dependent positron annihilation experiments have shown that clusters formed during initial NA are partially shallow positron traps, i.e. notably trap only far below 20 °C, but grow to deep traps in the course of NA [9,24]. Therefore, the initial state after quenching contains quenched-in vacancies and some initial (shallow and deep trap) clusters, which determines what happens during ensuing AA.

In all the alloys, even the shortest AA reduces $\tau_{1C}$ markedly. According to Figure 7, this indicates the loss of vacancies that are present in high numbers in the as-quenched state and possibly the formation of some clusters. For the shortest AA times, the initial values of $\tau_{1C}$ vary a lot and range from 163 ps to 222 ps depending on the alloy and heating conditions. $\tau_{1C}$ reaches a minimum value in all but one case (alloy 6-8, LM), after which it increases again to a final value up to 235 ps. This increase is associated with the formation of positron traps in which positrons live longer, see Figure 7. In the following, we shall discuss the measured data first for "*short AA*" that leads from the high $\tau_{1C}$ after quenching to a minimum value, then for "*intermediate AA*" that features a re-increase of $\tau_{1C}$ to 210 to 215 ps and finally for "*long AA*" that is characterised by a very pronounced further increase of $\tau_{1C}$. The main emphasis of this paper lies on short AA since this is the least explored regime and positrons are especially sensitive here.

### A. Short AA times

#### a. Interpretation of experiments

The most pronounced decrease of $\tau_{1C}$ is observed in alloy 4-4 dipped into LM for just 1 to 5 s. Low values just slightly above the value for defect-free Al (160 ps) indicate that most vacancies have been eliminated and the number of clusters that might have formed is limited or they do not trap positrons at 20 °C. The positron lifetime spectrum exhibits two distinct components (Figure 2b,c) with $I_{2+3}$~50%. Technically, this means that roughly one half of all positrons annihilate in the bulk because they do not encounter a vacancy or cluster on their diffusion paths through the alloy.

If all positron traps were vacancies, the two-state trapping model [11] could tell us the fraction of vacancies. Figure 8 (black dashed-dotted lines) shows that the lowest measured value for $\tau_{1C} \approx \bar{\tau}$ of 163 ps in Figure 2a corresponds to a site fraction of vacancy-related defects of $5 \times 10^{-7}$ (solid triangle). This can be compared to the fraction directly after quenching, $7.5 \times 10^{-5}$ (open triangle), following from an average positron lifetime of 230 ps estimated for that case (see comments in supplementary Sec. S3).

These values correspond to $4.5 \times 10^{-5}$ measured for a Al-1.1at.%Cu-1.7 at.%-Mg alloy after quenching and $1.2 \times 10^{-5}$ after additional AA. In Mg-free Al-Cu alloy values were much lower, pointing at Mg stabilising vacancies in complexes [25]. A similar action of Mg can be assumed for Al-Mg-Si alloys.

As the scatter of two-component decompositions can be considerable we have averaged all the values up to 100 min of AA. From Figure 2b we know that $\tau_{2+3}$ is on average 216 ps after short AA and not 245 ps as the two-state trapping model suggests (red dash-dotted line in Figure 8). This average agrees very well with a single value measured with the high-resolution spectrometer P and shows that most positrons are not trapped in vacancies but in clusters. As experiments at B and P were carried out at different temperatures this points at the predominance of clusters that are deep, i.e. trap at 'room temperature', unlike the clusters created during short NA in another study that were partially shallow positron traps [9]. In Figure 8, we now add a constant cluster contribution big enough to bring down $\tau_{2+3}$ and apply the three-state trapping model (full lines). As the lifetime $\tau_{2+3}$=216 ps is the result of a mixed contribution of $\tau_2$=210 ps and $\tau_3$=245 ps, trapping in vacancies is even lower than assumed above, namely $2.5 \times 10^{-7}$ in the example given in Figure 8 (split triangle). This is just about 3 times higher than the equilibrium vacancy fraction at 180 °C, $8 \times 10^{-8}$. Thus, short AA in LM can eliminate most vacancies while producing just few clusters.

After slower heating in oil, MS, or on the heating plate, the positron traps are much more densely distributed and fewer positrons annihilate in the bulk as seen by the trap contribution $I_{2+3}$ that is now ~80% instead of just ~50% for fast heating, see Figure 2c. $\tau_{2+3}$ is similar to that in LM, see Figure 2b. Therefore, more clusters have formed during slower heating than during fast heating.

A problem encountered when comparing short AA at different heating rates is that of an ambiguous definition of ageing time. Figure 9 helps to compare the four different heating media by displaying the effective time spent above 170 °C for each measurement. In Figure 2a, for example, the two-sided arrow compares AA in LM for 5 s and that in oil for 10 s. For heating in LM or oil, about 0.9 s or 6 s, respectively, are needed to reach 170 °C and therefore both samples spent about 4 s above 170 °C as marked by the corresponding arrow in Figure 9.

The scatter of experimental values of $\tau_{1C}$ (and therefore also $\bar{\tau}$) in Figure 2a and especially values much higher than 163 ps are thought to be the consequence of slightly varying conditions inflicted by the manually conducted transfer from the ice water after quenching to the LM.

b.  **Simulation of vacancy dynamics**

The thermo-kinetic software *MatCalc* allows us to simulate the loss of vacancies during complex heat treatment sequences [26]. The underlying physics is the FSAK model for vacancy loss [7]. In our case, vacancies are predominantly lost at dislocation jogs. We model vacancy losses using the parameters suggested by Ref. [27], apply a temperature profile close to our experiment and obtain Figure 10. A very high quenching rate (5000 K·s$^{-1}$) is required to obtain a post-quenching (after steps 1 and 2) vacancy fraction of $7.5 \times 10^{-5}$ high enough to explain the observed positron lifetime of 230 ps (Figure 8). A quench rate of 900 K·s$^{-1}$ in water at 20 °C has been reported for 2.5-mm thick samples [28] so that quenching of our just 1-mm thick samples in ice water could be that fast. Subsequent heating to 180 °C (step 3) leads to a decrease of vacancy fraction. Accepting the estimate of Figure 8 that the vacancy fraction is 2.5 × 10$^{-7}$ after AA in LM (dotted horizontal line), we obtain a time of ~40 s for this reduction by *MatCalc* calculation without vacancy-solute interactions, ~55 s with them included. Additional vacancy-cluster interactions would further slow down vacancy losses, but we are presently not able to simulate them (read below). Increasing the dislocation density (or the jog fraction) by a factor of 10 or even 100 speeds up vacancy loss and the latter value would explain the experimentally found fast vacancy loss in just ~1 s. Such high dislocation densities have been reported for quenched Al-10Mg alloys (5 × 10$^{13}$ m$^{-2}$) [29], Al-1.2Si alloys and in pure Al following quenching (2.2 × 10$^{13}$ - 10$^{14}$ m$^{-2}$, derived from [30-32]). However, the vacancy fraction after quenching would also be lowered by such an increase of vacancy sink fraction and would go to values incompatible with the high measured positron lifetimes after quenching. As a way out, dislocations could be thought to be created during quenching due to stresses caused by thermal gradients. If dislocations were progressively created predominantly towards the end of quenching at low temperatures a high vacancy fraction would be preserved.

Alternatively or in addition, vacancies could quickly form vacancy clusters during AA. They would have an only small effect on positron lifetime due to their low number but significantly lower the fraction of free vacancies and therefore contribute to the fast decrease of $\tau_{1C}$.

c. **Explanation of heating rate effect**

In Figure 10, the heating rate to 180 °C does hardly influence the vacancy loss dynamics. This is because cluster formation is not taken into account. Solutes are transported to emerging clusters mediated by vacancies, but the emerging clusters then interfere strongly with vacancies [33]. This feedback situation is difficult to model without knowing the relevant interactions, which is why we can discuss this in a qualitative way only.

After solutionising at 540 °C and subsequent fast quenching, excess vacancies partition very quickly into free vacancies, $x_v^{\text{free}}$, and vacancies bound to solute atoms, $x_v^{\text{bound}}$. The fraction $\eta$ of free vacancies (of all vacancies, $x_v^{\text{tot}}$) can be calculated by a thermodynamic model [34] that has been successfully applied to Sn solute atoms in Ref. [35]:

$$\eta = \frac{x_v^{\text{free}}}{x_v^{\text{tot}}}(T) = \frac{1+12x_s \exp(E/RT_{\text{SHT}})}{1+12x_s \exp(E/RT)}, \quad \text{with } x_v^{\text{free}} + x_v^{\text{bound}} = x_v^{\text{tot}}, \tag{1}$$

which has been evaluated in Figure 11 for our case. Without any vacancy losses one could reversibly move up and down temperature and the vacancies would repartition accordingly into free vacancies and bound vacancy-solute complexes. Only bound (but still mobile) vacancies, fraction (1-$\eta$), contribute to solute diffusion because they allow solute atoms to chance lattice sites. Whenever a bound solute-vacancy pair meets other solutes or clusters, it might stick to those, after which the vacancy moves away again after a residence time [36,37]. The sticking probability $p_{\text{stick}}(T)$ will be higher for lower temperatures as it is governed by the (poorly known) binding energy between a solute and a cluster. We write for the rate of cluster volume growth $\dot{V}_c$:

$$\dot{V}_c = \Gamma_v(T) \underbrace{(1-\eta(T)) x_v^{\text{tot}}}_{x_v^{\text{bound}}} p_{\text{stick}}(T), \tag{2}$$

where $\Gamma_v(T)$ is the jump frequency of a vacancy or solute-vacancy complex, which markedly increases with temperature. In the presence of vacancy sinks, the free vacancy site fraction will decrease as shown in Figure 10. During early ageing the rate of vacancy loss can be simplified to

$$\dot{x}_v = -\Gamma_v(T) \underbrace{\eta(T) x_v^{\text{tot}}}_{x_v^{\text{free}}} x_{\text{sink}}, \tag{3}$$

where $x_\text{sink}$ is the temperature-independent vacancy sink (essentially dislocation jog) fraction. Combining Eqs. (2) and (3) we obtain the ratio of clustering vs. vacancy loss:

$$\frac{\dot{V}_C}{|\dot{x}_v|} = \frac{1-\eta(T)}{\eta(T)} \frac{p_\text{stick}(T)}{x_\text{sink}}. \qquad (4)$$

This quantity is higher for 20 °C than for 180 °C as both factors decrease with T (dashed line in Figure 11 and the above mentioned behaviour of $p_\text{stick}$). Thus, during NA many clusters are formed while vacancies are lost slowly and the positron lifetime drops from an initially high value to a typical cluster value (i.e. ~215 ps, see supplementary Fig. S1a). During instant AA (fictitious infinitely high heating rate to 180 °C), best represented by AA in LM, fewer clusters are formed and more vacancies are lost, Eq. (4), and therefore the positron lifetime decreases to values much below 215 ps, see Figure 2a and Figure 3. Of course this happens many orders of magnitude faster during AA than during NA.

Eq. (4) is valid for the initial ageing stage. Once clusters have been formed the picture changes. Clusters trap vacancies much more efficiently than solute atoms. According to Ref. [36], the binding energy in Eq. (1) should be replaced by $N \times E$ for $N$ equal atoms in a cluster. Figure 11 shows how low the fraction of free vacancies for $N = 9$ gets. Moreover, vacancies are trapped in clusters for a longer time [36]. Vacancies in clusters can induce internal ordering processes and displace clusters. Thus, they contribute to further clustering while reducing vacancy losses, i.e. the ratio in Eq. (4) is further increased. This explains qualitatively the observations during slow heating to 180 °C: During the few seconds at lower temperatures, 20 °C and above, clustering sets in, while relatively few vacancies are lost. At higher temperatures and finally at 180 °C, the clustering rate is larger than after instant heating because of the larger ratio in Eq. (4).

In the above considerations attractive solute-vacancy and cluster-vacancy interactions have been assumed although claims of repulsive interactions, e.g. between Mg and vacancies exist [38] and much of the literature data – especially older literature – is not consistent (see overview in Ref. [39]). In the light of the latest literature, most next neighbour interactions are indeed attractive, while on some more distant shell interactions can turn to repulsive [40-42].

### d. Further experiments for short AA times

Comparing different alloys provides more insight into the phenomena during short AA. Figure 4 compares values for $\tau_{1C}$ after 1 s of AA in LM for 6 alloys as a function of solute content (solid symbols). Higher solute content clearly causes a higher $\tau_{1C}$, implying that fewer vacancies are lost and/or more clusters are formed under the same conditions. This finding is also reflected by a two-component decomposition of positron lifetime that was possible in one case for alloy 6-8 (first red point in Figure 3). Here, 84% of all positrons are trapped by clusters or vacancies compared to less than 50% in alloy 4-4. In the literature, a similar value of ~90% is given, however after even slower heating [43]. Therefore, alloy 6-8 contains more vacancies and clusters than alloy 4-4 after a similar short AA treatment. There are two reasons for this: firstly, vacancies are more likely to interact with solute atoms and to form clusters when the solute content is high, which helps to retard vacancy annihilation. The first term in Eq. (4) is higher in alloy 6-8, see Figure 11. Secondly, as discussed above, some clusters might have formed already during quenching and such formation is expected to be more pronounced for higher solute contents [9,44]. Consequently, even after heating up alloy 6-8 at the fastest rate (0.3 s at 300 °C in LM) $\tau_{1C}$ is still 190 ps, much higher than the 163 ps measured in alloy 4-4 after 1 s in LM at 180 °C.

If such pre-existing clusters in alloy 6-8 have this effect, clusters formed in alloy 4-4 by carrying out some natural ageing before AA should have the same effect. This is actually true as shown in Figure 4 (red symbols) where up to 80 min of natural pre-ageing (NPA) in alloy 4-4 lead to a similar effect as increasing the solute content.

Finally, we look at changes of electrical resistivity caused by 10 s of AA, Figure 6. AA is expected to change resistivity in 3 ways: (i) any loss of vacancies reduces resistivity with an estimated rate of 1.9 µΩ·cm/at.% [45] (not known with great precision [46,47]). Taking the vacancy fraction of $7.5 \times 10^{-5}$ used above for the quenched state, the reduction would be 0.014 µΩ·cm for a loss of all vacancies, which is 2.1% or 1.4% of the absolute resistivity given in Figure 6 for our alloys 4-4 or 6-8, respectively. (ii) Clustering is known to increase resistivity [48]. The maximum increase during very long NA is ~6% for alloy 6-8 [13,15]. (iii) Solute depletion from the matrix eventually decreases resistivity, but this contribution should be small after 10 s of AA. In alloy 4-4 aged in LM, the resistivity reduction by 1% is therefore compatible with the loss of most vacancies and some cluster formation that does only partially compensate the effect of vacancy loss. In alloy 6-8 heated in oil, in contrast, the increase by clustering is stronger than the reduction by vacancy loss, hence a net increase. As electrical resistivity and positron lifetime are influenced by vacancy loss (reducing both) and clustering (increasing both) in the same way, the two observables are correlated, see Figure 6.

## B. Intermediate AA times

After short ageing has reduced $\tau_{1C}$ in alloy 4-4, $\tau_{1C}$ increases again in intermediate ageing stages (Figure 2a). In alloy 6-8, the minimum is very shallow and $\tau_{1C}$ is almost constant (Figure 3). For all the alloys and heating conditions, 100 – 200 min of AA lead to $\tau_{1C}$ values between 208 ps and 219 ps, pointing at largely saturated positron trapping into the precipitates typically formed during AA.

The stage of constant $\tau_{1C}$ for alloy 6-8 corresponds to the highest increase of hardness, see Figure 5. Precipitates grow from GP zones to β″ phase here but $\tau_{1C}$ changes little because positron trapping is already saturated. In alloy 4-4, however, $\tau_{1C}$ increases a lot up to 200 min AA, but hardness changes only marginally. Here, $\tau_{1C}$ increases so much because at the positron lifetime minimum, trapping into clusters is not saturated and forming small but easily shearable precipitates with a positron lifetime of $\tau_{2+3}$ ~216 ps (Figure 2b) has a large effect on $\tau_{1C}$ but not on hardness.

Studies in the literature on similar alloys show a comparable behaviour [22,43]. Altogether, these ageing stages are not very interesting from the viewpoint of positron annihilation since the lifetime values obtained are rather insensitive to the microscopic configuration. The slight ups and downs observed in the $\tau_{1C}$ curves are hard to interpret without excessive speculation.

### C. Long AA times

AA for ≥200 min leads to an increase of $\tau_{1C}$ to values above 220 ps and eventually to 235 ps for both alloys and all heating conditions. Figure 2b shows that $\tau_{2+3}$ for alloy 4-4 increases to 230 ps – 240 ps for long AA times. Obviously, the character of the positron traps changes upon AA. A vacancy-related trap cannot be the reason as all vacancies are in thermal equilibrium after long AA (Figure 10). In accordance with Refs. [23,43], we assume that additional open volume associated with the partial incoherency of the precipitates formed at this stage causes the increase of $\tau_{2+3}$. As the precipitation sequence involves a transition from early GP zones to β" and/or β', the degree of lattice mismatch and lack of coherency increases in this order. β' formation is associated with a global shrinkage of a sample due to the removal of Mg from the matrix that overcompensates the expansion of β' [49,50] and tensile stresses around individual β' precipitates build up (other than for β" where the stresses are compressive). This facilitates the formation of open volume. As $I_{2+3}$ remains at ~80% but $\tau_{2+3}$ increases, $\tau_{1C}$ is notably increased to almost 235 ps (Figure 2b). The increase observed is much more pronounced in this study than in other work [22,23]. Only Resch et al. [43] find a comparable increase to 231 ps after 2 weeks of AA at 180 °C. Over-ageing for 1 week at 180 °C lets precipitates coarsen and hardness correspondingly decreases but the value for $\tau_{1C}$ remains high. Apparently even the coarsened precipitates trap all positrons efficiently and prevent the appearance of a bulk component that would reduce $\tau_{1C}$ again. The reason for this lies in the extended nature of the precipitates and the large value for the specific trapping rate of their interfaces [51] as can be calculated using the formalism given in Ref. [52].

The increase of $\tau_{1C}$ actually takes place at a later time in alloy 6-8 than in alloy 4-4 (blue arrow in Figure 5). As ageing proceeds faster in alloy 6-8 than in 4-4 (expressed by faster hardening, green arrow) this requires attention. The fast $\tau_{1C}$ increase coincides with fast hardening in alloy 4-4 as the black dash-dotted line shows but not in alloy 6-8. One possibility is that in alloy 4-4 hardening is mainly driven by formation of β′ phase and not of β″. The already mentioned long positron lifetime in β′ would then cause the rapid increase. In fact, a recent transmission electron microscopic study has recently shown for an alloy similar to our alloy 4-4 that β′ forms directly during AA without the need to first form β″ [53].

## III. CONCLUSIONS

Artificial ageing (AA) of solutionised and quenched Al-Mg-Si alloys of different compositions can decrease the one-component positron lifetime $\tau_{1C}$ by up to 80 ps in just a few seconds.

In the lean alloy 4-4 heated to 180 °C within 1 s, vacancies diffuse to sinks quickly and do hardly assist in clustering. The state produced is that of a sparse population of clusters and some residual excess vacancies. Therefore, positrons annihilate with a low lifetime in the bulk lattice since most of them are not trapped. The fast annihilation observed points at a large site fraction of vacancy sinks that might be formed during quenching. Vacancy clustering could also contribute to the decrease of positron lifetime.

The positron lifetime after AA is always longer and hence the trap density higher when clusters or precipitates are formed before or during AA, thereby limit further vacancy losses, which in turn enables further clustering - a self-amplifying effect. This happens whenever:

- The alloy contains more solute: in this case, more clusters are formed already during quenching after solutionising.
- The alloy is heated slowly to 180 °C: in this case, clustering sets in at lower temperatures, which then delays losses of vacancies at higher temperatures.
- The alloy is naturally pre-aged before AA: the clusters formed act in the way described above.

Only because positron lifetime is so sensitive to minimal changes of open volume in crystalline systems these phenomena give rise to such pronounced positron lifetime changes. Electrical resistivity as the second most sensitive method yields only weak signals, hardness and DSC measurements none.

**Acknowledgements**

We would like to thank Jürgen Hirsch from Hydro Aluminium Bonn for providing the alloys and the Deutsche Forschungsgemeinschaft for funding (Ba 1170/22). Qianning Guo and Xingpu Zhang thank the China Scholarship Council (CSC) for research fellowships. We are grateful to Roland Würschum for pointing out that the formalism given Ref. [52] can be used to show that saturated positron trapping can take place even in overaged alloys.

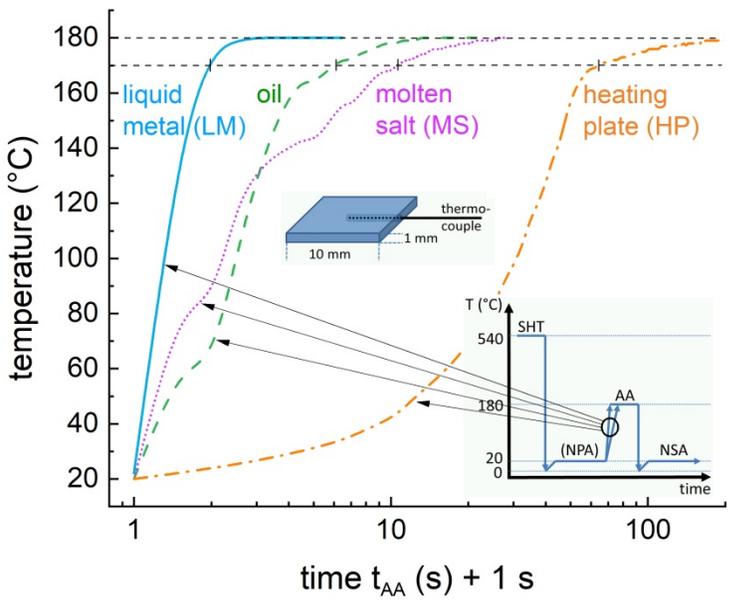

Figure 1. Heating profiles applying four different media for AA at 180 °C. The starting time is arbitrarily set to 1 s to be able to use a logarithmic time scale. Note that the linear heating ramp of the heating plate appears as an exponential on this scale. The heating curve in molten salt shows some kinks as upon immersion of a cold sample some salt first solidifies and then melts after a short time. In oil, a deviation of slope is observed around 70 °C for an unknown reason. Insets: setup of temperature measurement and general heating program of experiments.

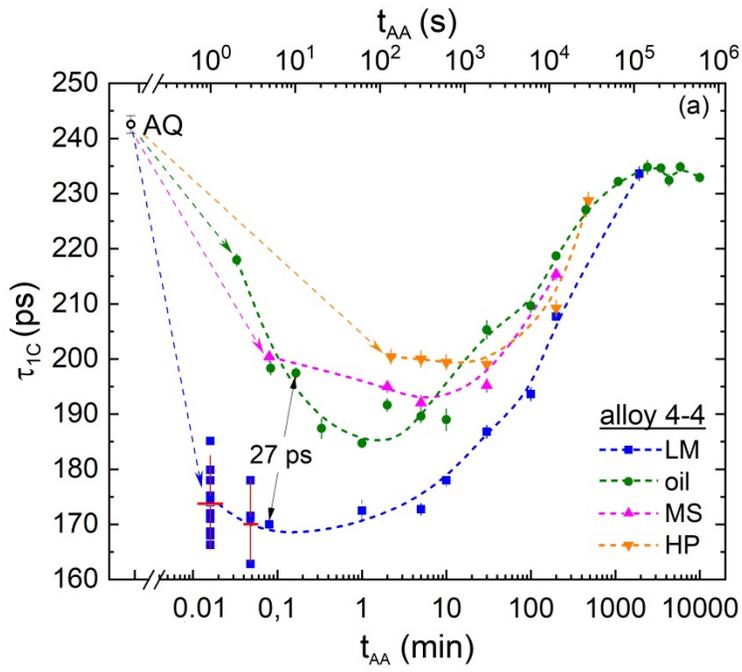
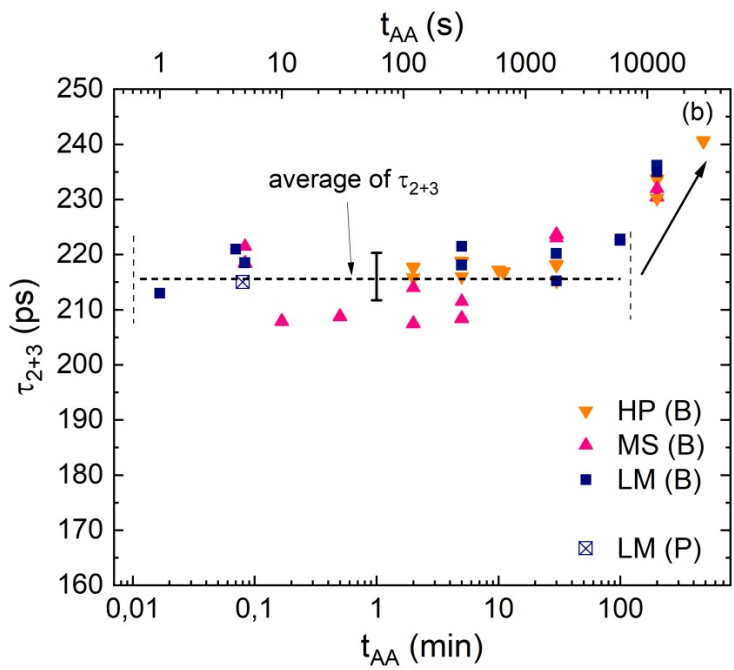

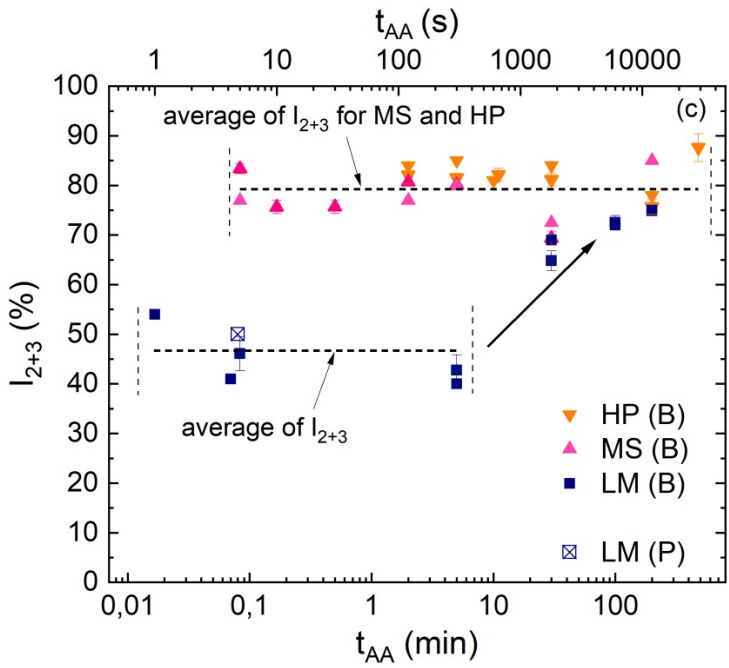

Figure 2. Evolution of positron lifetime $\tau_{1C}$ in alloy 4-4. (a) $\tau_{1C}$ as a function of AA time in different heating media. Error bars refer to individual fitting uncertainty except for the red bars, where standard deviation of averages of different measurements and uncertainty in AA time are expressed. Double sided arrow corresponds to Figure 9. (b,c) Trap lifetimes $\tau_{2+3}$ and intensities $I_{2+3}$ as derived from positron lifetime decompositions for alloy 4-4. Different symbols specify the AA medium used. All experiments were carried out at ~20 °C ('B') except for the one represented by crossed squares that was measured in high resolution at −40 °C ('P'). The broken lines indicate averages of regimes considered approximately constant. Arrows mark the trends discussed in the text.

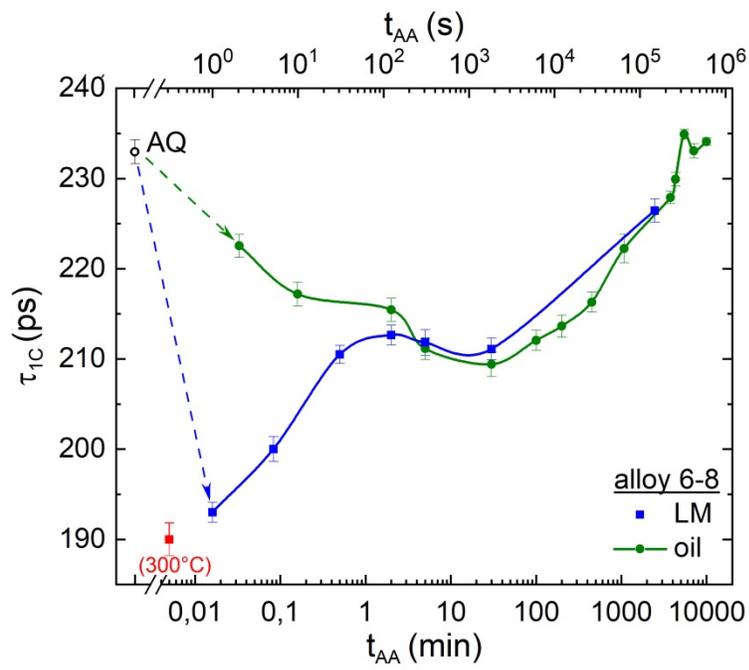

Figure 3. Evolution of the one-component positron lifetime $\tau_{1C}$ in alloy 6-8 as a function of AA time at 180 °C in different heating media. The red symbol corresponds to 0.3 s annealing at 300 °C.

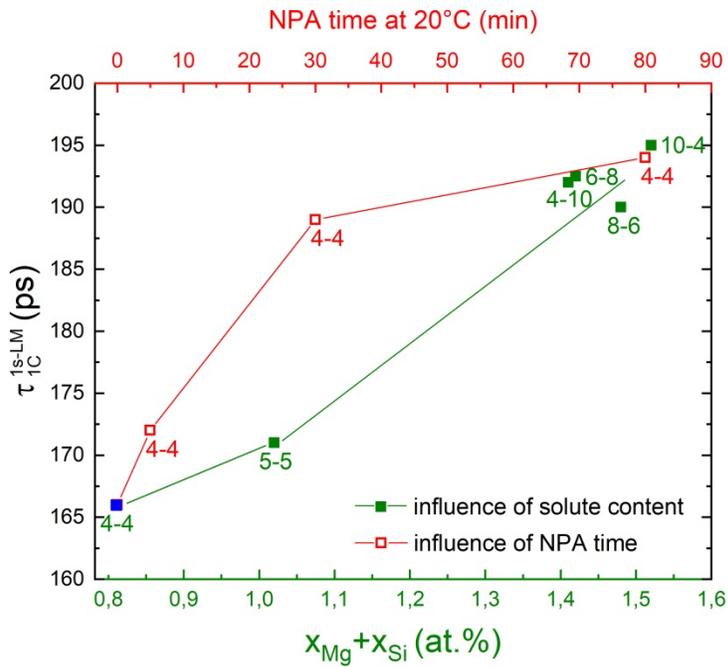

Figure 4. $\tau_{1C}$ after 1 s of AA in LM for various alloys with different Mg and Si contents (full symbols). Data for alloys 4-4 and 6-8 is the same as in Figure 2a (lowest value chosen) and Figure 3. For alloy 4-4, an additional NPA step (5, 30 or 80 min) has been included before AA (open symbols).

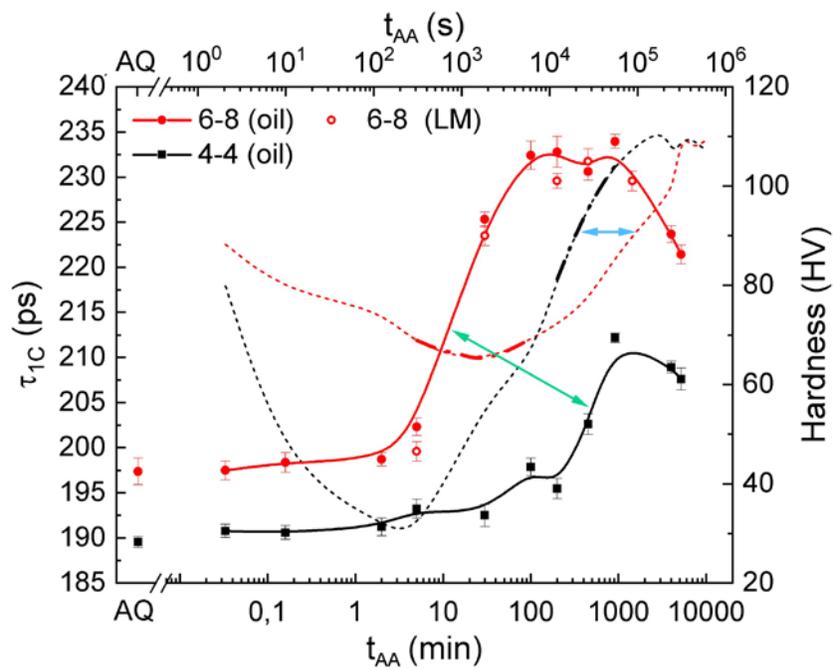

Figure 5. Hardening of alloy 4-4 and 6-8 at 180 °C in oil (and in LM for some cases). The thin broken lines represent the positron lifetimes $\tau_{1C}$ taken from Figure 2a and Figure 3. On these lines, the dash-dotted sections emphasise the regions of main hardening. The double-sided arrows show the offset between main hardening and final increase of $\tau_{1C}$ for the two alloys.

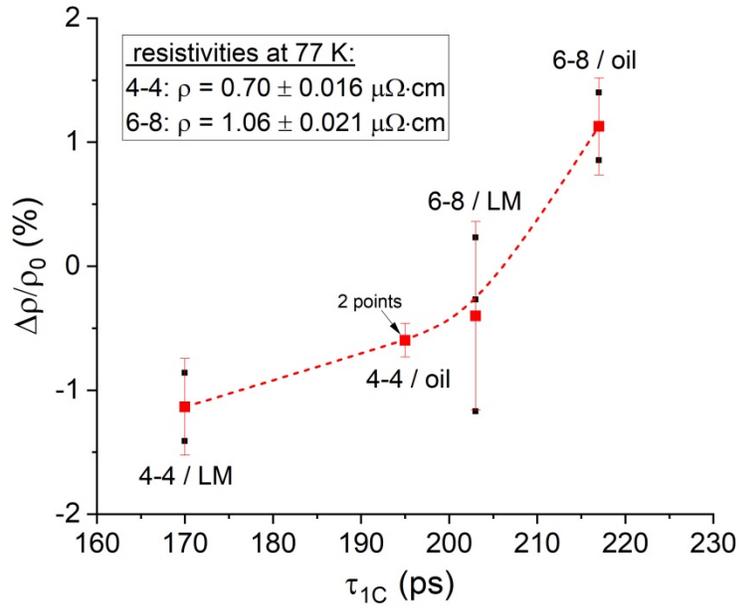

Figure 6. Relative change of electrical resistivity during AA for 10 s measured for two alloys and two media (LM, oil). Black points are individual measurements, red points are averages. The data is related to positron lifetimes derived from Figure 2a and Figure 3.

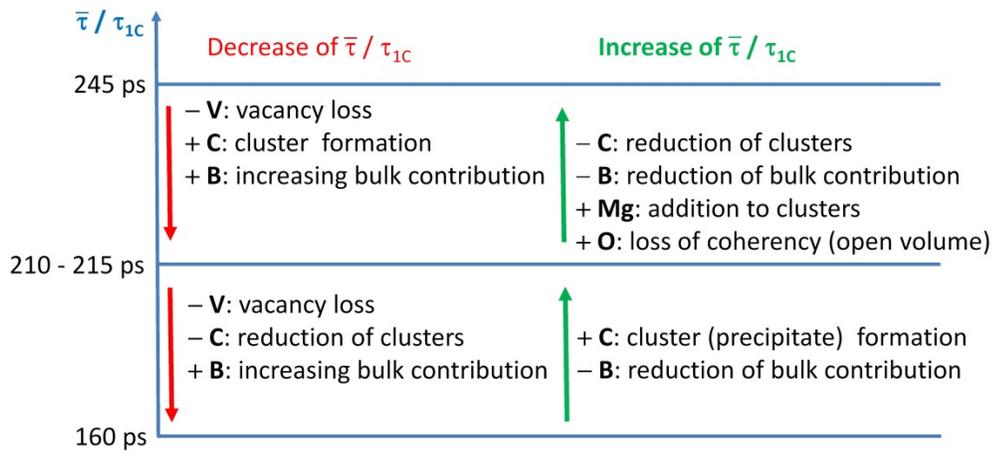

Figure 7. Alternative mechanisms governing the change of the averaged ($\bar{\tau}$) or one-component positron lifetime ($\tau_{1C}$) during ageing. '+' and '−' denote an increase or decrease of a specific contribution. In this paper, vacancy (V) loss, clustering (C), loss of coherency (O) and the change of bulk annihilation (B) is discussed. Change of Mg content (Mg) governs natural ageing [8] and is not relevant here.

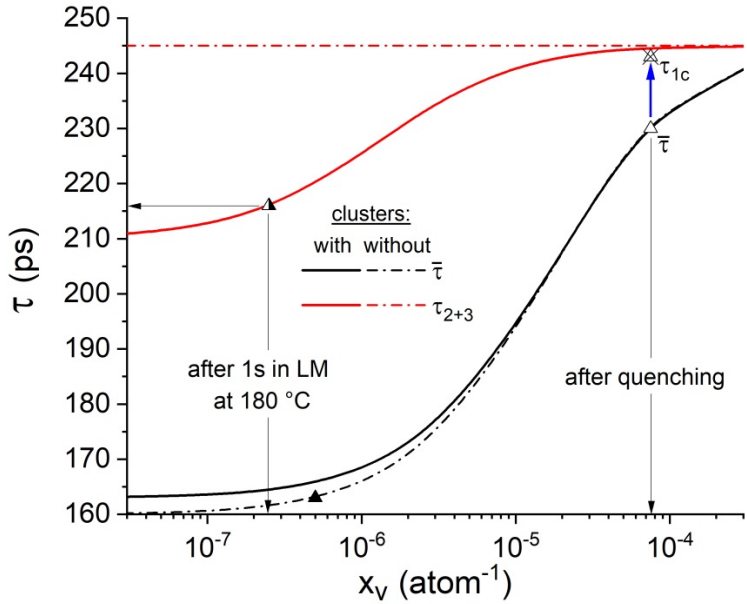

Figure 8. Relationship between average ($\bar{\tau}$) or trap-related ($\tau_{2+3}$) positron lifetimes and the site fraction of vacancies $x_v$ as calculated from the trapping model defined in Ref. [11] and using a bulk lifetime of $\tau_b$=160 ps, a cluster contribution given by $\tau_c$ = 210 ps and a trapping rate $\kappa_c$ = 4×10$^{-4}$ ps$^{-1}$, and a vacancy-related component $\tau_v$= 245 ps with a trapping rate $\kappa_c = \lambda_v x_v$, with $\lambda_v$=430 ps$^{-1}$ taken from Ref. [54]. The average positron lifetime $\bar{\tau}$ is in general reflected by the measured $\tau_{1C}$ rather well (see supplementary Fig. S2). For very high $x_v$, however, $\tau_{1C}$ might overestimate $\bar{\tau}$, which is just 230 ps (open triangle) and not 243 ps (crossed triangle, further discussion, see end of supplementary Sec. S3). The broken lines represent calculated positron lifetimes in the presence of just vacancies, the solid ones for an additional constant cluster component. Triangles mark special points discussed in the text.

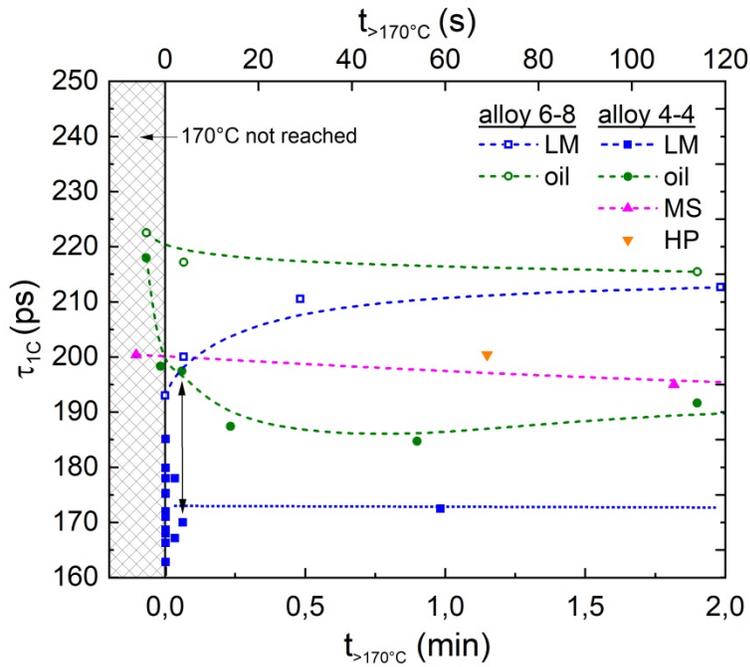

Figure 9. Positron lifetimes $\tau_{1C}$ measured in alloys 4-4 and 6-8 after AA in various media for up to 2 min. Data is from Figure 2a and Figure 3, but time axis is linear and corrected for the time the alloy has spent at temperatures below 170 °C, i.e. 0.9 s, 6 s, 11 s or 63 s were subtracted from the times in Figure 2a and Figure 3 for heating in LM, oil, MS and heating plate, respectively. Negative values of t (hatched area) indicate that a sample did not reach 170 °C but will still need time |t| to reach 170 °C. The vertical arrow compares two experiments in LM where 4 s were spent above 170 °C.

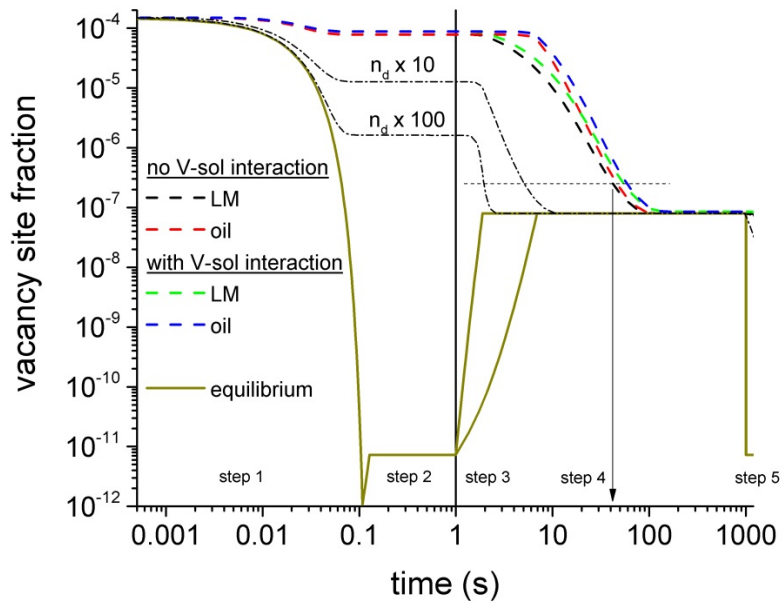

Figure 10. *MatCalc* simulation of non-equilibrium vacancy site fraction in alloy 4-4 exposed to a temperature profile comprising 5 steps, namely 1) quenching from 540 °C to 0 °C at −5000 K·s$^{-1}$; 2) brief holding at 20 °C (time irrelevant) up to t=1 s; 3) heating to 180 °C at two different rates (170 K·s$^{-1}$ and 25.5 K·s$^{-1}$ approximating LM and oil); 4) holding at 180 °C and 5) quench to 20 °C, after which positron lifetime is measured. The olive line is the equilibrium vacancy site fraction. The parameters chosen are: dislocation density of $n_d$ =3 × 10$^{11}$ m$^{-2}$ and jog fraction 0.02 (i.e. one jog every 50 atoms), leading to a jog site fraction of 2.5 × 10$^{-10}$. The grain size was 0.5 mm [8]. Vacancy-solute interaction energies were the same as used in Ref. [55]. The dash-dotted curves are variants of the broken black curve with 10 or 100 times higher dislocation densities. All values have been multiplied by the entropy factor $e^{-S/k}$=2 not included in MatCalc, the value of which is from Ref. [45].

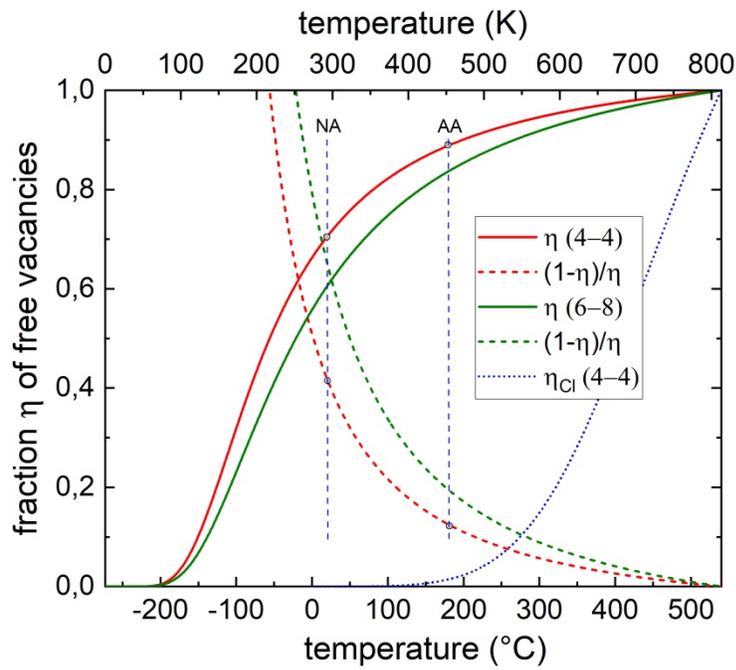

Figure 11. Fraction $\eta$ of the vacancies quenched-in from $T_{SHT}$=540 °C that are free as calculated from the thermodynamic model of [34,35] given by Eq. (1), solid lines. A binding energy $E$ = 50 meV between a solute atom and a vacancy is assumed and $x_s$=8 × $10^{-3}$ or 1.4 × $10^{-2}$ as for alloys 4-4 or 6-8, respectively. The dashed lines represent the factor $(1-\eta)/\eta$ occuring in Eq. (4). The exemplary curve for a cluster containing 9 atoms is based on $E$ = 450 meV, $x_s$=3 × $10^{-5}$ and obtained by replacing 12 by 37 neigbours in Eq. (1).

Table 1. Compositions of alloys as determined by optical emission spectroscopy and given in wt.% (at.%).

| Alloy designation | Mg | Si |
|---|---|---|
| 4-4 | 0.39 (0.43) | 0.40 (0.38) |
| 5-5 | 0.46 (0.51) | 0.54 (0.52) |
| 4-10 | 0.40 (0.44) | 1.01 (0.97) |
| 6-8 | 0.59 (0.66) | 0.79 (0.76) |
| 8-6 | 0.80 (0.89) | 0.61 (0.59) |
| 10-4 | 1.02 (1.13) | 0.41 (0.39) |

---

**†** It was attempted to show a difference in the DSC signal obtained on samples after 10 s AA in LM and oil following the standard protocol give in Ref. [14] While as-quenched samples feature a pronounced clustering and dissolution signal as described before [15], 10 s AA removes this peak and the corresponding dissolution trough almost completely. AA in oil and LM give rise to a very similar signal. Hence, DSC is not sensitive enough to detect the differences seen by positron annihilation and resistivity measurement.